\documentclass[twocolumn,showpacs,preprintnumbers,amsmath,amssymb]{revtex4}

\usepackage{graphicx}
\usepackage{dcolumn}
\usepackage{bm}

\begin{document}


\title{Phase transitions in diluted  negative-weight percolation models}

\author{L. Apolo$^{1,2}$}
\email{lapolo00@ccny.cuny.edu}
\author{O. Melchert$^2$}
\email{melchert@theorie.physik.uni-oldenburg.de}
\author{A. K. Hartmann$^2$}
\email{alexander.hartmann@uni-oldenburg.de}
\affiliation{
$^1$ City College of the City University of New York, New York, New York 10031, USA\\
$^2$ Institut f\"ur Physik, Universit\"at Oldenburg, 
Carl-von-Ossietzky Strasse, 26111 Oldenburg, Germany
}

\date{\today}


\begin{abstract}
We investigate the geometric properties of loops on two-dimensional lattice
graphs, where edge weights are drawn from a distribution that allows
for positive and negative weights. 
We are interested in the
appearance of spanning loops of total negative weight.  The resulting
percolation problem is fundamentally different from conventional
percolation, as we have seen in a previous study 
of this model for the undiluted case.

 Here, we investigate how the percolation transition
 is  affected by
additional dilution. We consider two types of dilution: 
either a certain fraction of edges exhibit zero weight, or a fraction
of  edges is even absent.
We study these systems numerically using exact combinatorial
optimization techniques based on suitable transformations of the
graphs and applying matching algorithms.
We perform a finite-size scaling analysis
to obtain the phase diagram and determine the critical properties
of the phase boundary.
 We find that the first type of dilution does
not change the universality class compared to the undiluted case
 whereas the second type of
dilution leads to a change of the  universality class.
\end{abstract} 

\maketitle

\section{Introduction \label{sect:introduction}}

Percolation is one of the most fundamental problems in statistical
mechanics \cite{stauffer1979,stauffer1994}. Its ideas apply to
physical disciplines as different as  solid-state physics and
cosmology and it has the ability to describe many phase transitions in
a completely geometric and probabilistic context.  The pivotal
question in percolation theory is that of connectivity. Consider 
for example
random-bond percolation where one studies a lattice that is randomly
filled with a certain fraction  of edges. Clusters of connected sites
are then analyzed regarding their geometric properties.  Depending on
the fraction of edges  on the lattice, the geometric properties of the
clusters change, leading from a phase with rather small and
disconnected clusters to a phase where there is basically one  large
cluster dominating the lattice.  Therein, the appearance of an
infinite, i.e.\ percolating, cluster is described by 
a second order phase
transition.  In the past decades, a large number of percolation
problems have been investigated in various contexts
through numerical simulations. Among those are  problems
where the objects of interest are string-like, rather than clusters as
in the standard case explained above. 
The geometric properties of string-like objects have
been studied and quantified in diverse contexts
\cite{schwartz1998,hindmarsch1995,engels1996,schakel2001,bittner2005,wenzel2005,antunes1998,pfeiffer2003}.
Recently \cite{melchert2008} we have investigated {\em negative weight
percolation} (NWP), a problem in similar vein but with subtle
differences compared to other string-like percolation problems.  In
NWP, we consider a regular lattice with periodic boundary
conditions  where all adjacent sites are joined by undirected
edges. Weights are assigned to the edges, representing quenched random
variables drawn from a distribution that allows for edge weights of
either sign.  The details of the weight distribution is therein
controlled by a disorder parameter.  For a given  realization of the
disorder, we then ask for a configuration of loops, i.e.\ closed
polygons on the lattice, such that the sum of the edge weights that
make up the loops is minimal. The problem of finding these loops can
be cast into a minimum-weight-path (MWP) problem as will be explained
below in more detail. This MWP is also of relevance for  the problem
of finding domain-wall excitations in 2d Ising spin glasses
\cite{melchert2007}.  One feature of this particular MWP problem is
that there is no modularity, meaning that a path obeying the
optimization constraint  can have subpaths that do not meet the
optimization criterion. We further impose the constraint that the
loops are not allowed to intersect, as a result there is no definition
of clusters in NWP.  

Previously we investigated NWP on two-dimensional lattices. We found a
critical value of the disorder parameter above which percolating loops
emerge in  the limit of large system sizes. We further determined a
whole set of critical exponents  that characterize the underlying
disorder-induced and geometric phase transition.  These exponents
where clearly different from those describing other percolation
transitions.  Moreover, the critical exponents where found to be
universal in $2d$, meaning that they do not  depend on the details of
the disorder distribution or the lattice geometry.  

Albeit there are
essential differences between NWP and usual bond percolation described
above,  the features of the latter are important if we ask for the
effect of dilution on the  characteristics of NWP.  Here, we
investigate the effect of dilution on the critical properties of
NWP. Therefore we use observables from percolation theory and a 
finite-size-scaling (FSS) analysis to probe critical lines in the
disorder-dilution plane for two different  types of dilution described
below. Further, we check whether the critical exponents change along
the critical lines. We therefore analyze the critical exponent $\nu$
that describes the divergence of a typical length scale in NWP as we
approach the critical  point. For one distinguished point in the
disorder-dilution plane, a more complete analysis involving more
critical exponents is given as well.

The paper is organized as follows. 
In section \ref{sect:model}, we introduce the model in 
more detail and we outline the algorithm 
we used to obtain the loop configurations. In section \ref{sect:results}, we 
list the results of our numerical simulations followed by a summary and 
conclusions in section \ref{sect:conclusions}.

\begin{figure}[t]
\centerline{
\includegraphics[width=1.0\linewidth]{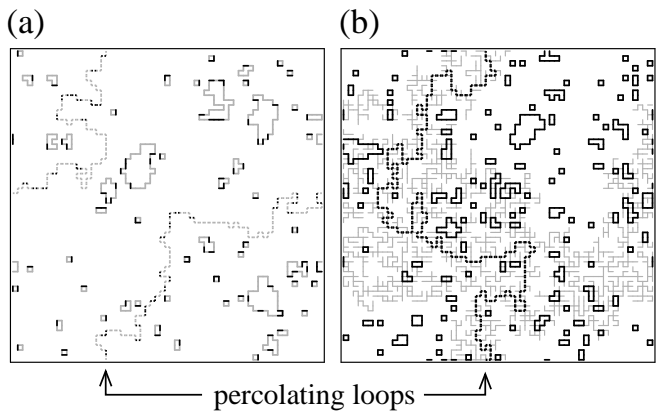}}
\caption{Samples of $L=64$ loop configurations for the two different
types of disorder investigated. Percolating loops appear as dashed 
lines and are marked by arrows.
(a) $\rho=0.1$, $p_{\rm{I}}=0.3940$, where $p_{\rm{I}}$ is the fraction of edges with
weight equal to zero. A loop can include those edges without
increasing its configurational energy. Edges with zero weight that 
are included in loops are colored grey.
(b) $\rho=1.0$, $p_{\rm{II}}=0.4998$, where $p_{\rm{II}}$ is the fraction of 
missing edges on the lattice. Those edges that comprise the
cluster of connected edges on which the percolating loop is located appear
as thin grey lines.
\label{fig1ab}}
\end{figure}  

\section{Model and Algorithm\label{sect:model}}
In the framework of this paper, we consider $2d$ square lattices
$G\!=\!(\mathcal{V},\mathcal{E})$  with side length $L$ and fully
periodic boundary conditions (BCs), i.e. $N\!=\!L\times L$ sites
$i\!\in\!\mathcal{V}$ and a maximal number of $2N$ (in case of no
dilution) edges  $\{i,j\}\!\in\!\mathcal{E}$ joining adjacent sites
$i,j\!\in\!\mathcal{V}$.  We further assign a weight/cost
$\omega_{ij}$ to each edge contained in $\mathcal{E}$, representing
quenched random variables that  introduce disorder to the lattice.
One realization of the undiluted disorder therein consists of a 
fraction $\rho$
of edge weights drawn from a Gaussian distribution with zero mean and
unit width, together with edges with unit weight 
(fraction $1-\rho$).  This
weight distribution explicitly allows for loops $\mathcal{L}$,
i.e. closed polygons on the lattice,  with a negative total weight
$\omega_{\mathcal{L}}\!=\!\sum_{\{i,j\}\in\mathcal{L}}\omega_{ij}$.
For any nonzero value of the disorder parameter $\rho$, a sufficiently
large lattice will exhibit at  least small loops with a weight smaller
than zero. If the disorder parameter is sufficiently large,
even large loops with negative weight will appear, which span the
system. Hence, the system is percolating in this case.
Here, as an additional source of disorder, we consider two
different types of dilution, affecting the lattice edges:

\begin{itemize}
 \item Type I:
The disorder is characterized by a fraction $p_{\rm{I}}$ of edge weights
$\omega_{ij}\!=\!0$.  This does not mean that the respective bonds are
absent in terms of bond percolation.  Instead, a loop can include
these edges without an increase of the configurational energy, see
figure \ref{fig1ab}(a); 

\item Type II: The lattice is diluted by a
fraction $p_{\rm{II}}$ of \emph{absent} bonds.  Depending on the value of
$p_{\rm{II}}$, this induces a usual random bond percolation process that
will affect the negative weight percolation (NWP) of loops, see figure
\ref{fig1ab}(b).
\end{itemize}

  Given $G$ together with a realization of the disorder,
we determine a set $\mathcal{C}$ of loops such that the
configurational energy, defined as the sum of all the loop-weights
$E\!=\!\sum_{\mathcal{L} \in \mathcal{C}} \omega_{\mathcal{L}}$, is
minimized.  As further optimization constraint, the loops are not
allowed to intersect and generally, the  weight of an individual loop
is smaller than zero. Note that $\mathcal{C}$ may also be empty.
Clearly, the configurational energy $E$ is the quantity subject to
optimization and the result of the optimization procedure is a set of
loops $\mathcal{C}$, obtained using an  appropriate transformation of
the original graph as detailed in \cite{ahuja1993}.  So as to identify
the edges that constitute the loops for a particular instance  of the
disorder, we need to obtain a minimum-weighted perfect matching (MWPM)
\cite{cook1999,opt-phys2001} on the transformed graph.

Here, we give a brief description of the algorithmic procedure that yields a 
minimum weight set of loops for a given realization of the disorder. 
Figure \ref{fig2abcd}
illustrates the 3 basic steps, detailed below:

\begin{figure}[t!]
\centerline{
\includegraphics[width=1.0\linewidth]{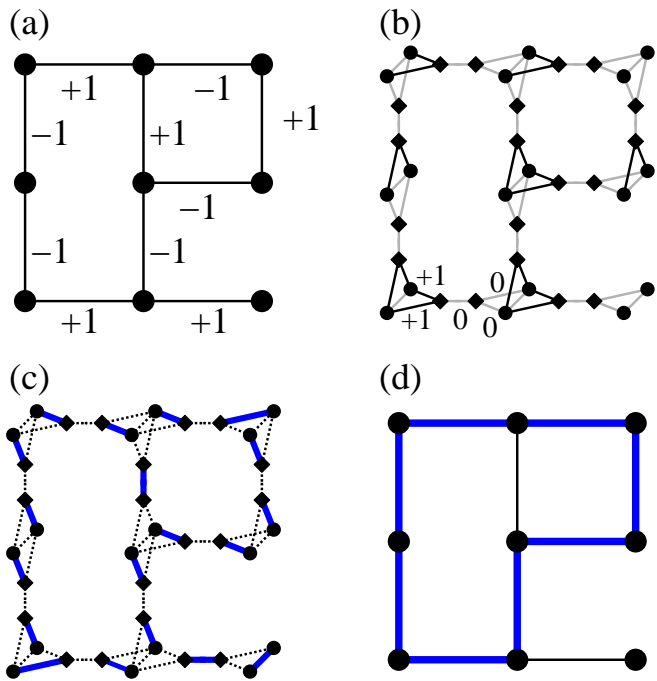}}
\caption{ Illustration of the algorithmic procedure.
(a) original lattice $G$ with edge weights, 
(b) auxiliary graph $G_{\rm A}$ with proper weight assignment: black 
edges carry the same weight as the respective edge in the original
graph and grey edges carry zero weight. 
(c) minimum-weight perfect matching $M$: bold edges are matched 
and dashed edges are unmatched.
(d) loop configuration (bold edges) that corresponds to the MWPM 
depicted in (c).
\label{fig2abcd}}
\end{figure}  

(1) each edge, joining adjacent sites on the original graph $G$,  is
replaced by a path of 3 edges.  Therefore, 2  ``additional'' sites
have to be introduced for each edge in $\mathcal{E}$.  Therein, one of
the two edges connecting an additional site to an original site gets
the same weight as the corresponding edge in $G$. The remaining  two
edges get zero weight.  The original sites $i\in \mathcal{V}$ are then
``duplicated'',  i.e. $i \rightarrow i_{1}, i_{2}$, along with all
their incident edges and the corresponding weights. 
 For each of these pairs of duplicated sites,
one additional  edge $\{i_1,i_2\}$ with zero weight is added that
connects the two sites $i_1$ and $i_2$.  The resulting auxiliary graph
$G_{{\rm A}}=(\mathcal{V}_{{\rm A}},\mathcal{E}_{{\rm A}})$  is shown
in figure  \ref{fig2abcd}(b), where additional sites appear as squares and
duplicated  sites as circles. Figure \ref{fig2abcd}(b) also illustrates
the weight  assignment on the transformed graph $G_{{\rm A}}$.  Note
that while the original graph (figure \ref{fig2abcd}(a)) is symmetric, the
transformed graph (figure \ref{fig2abcd}(b)) is not. This is due to the
details of the mapping procedure and the particular weight assignment
we have chosen.  A more extensive description of the mapping can be
found in \cite{melchert2007}.

(2) a MWPM on the auxiliary graph is determined via  exact
combinatorial optimization algorithms \cite{comment_cookrohe}.  A MWPM
is a minimum-weighted subset $M$ of $\mathcal{E}_{\rm A}$, such that
each site  contained in $\mathcal{V}_{\rm A}$ is met by precisely one
edge in $M$.  This is illustrated in figure \ref{fig2abcd}(c), where the
solid edges  represent $M$ for the given weight assignment. The dashed
edges are  not matched.  Due to construction, the auxiliary graph
consists of an even number of sites  and since there are no isolated
sites, it is guaranteed that a perfect matching exists.  Even if
theres no edge on $G$, i.e.\ $p_{\rm{II}}\!=\!1$ in terms of the dilution
introduced above. In that extreme case, $G_{{\rm A}}$ consists solely
of the ``duplicated'' sites that are joined pairwise by zero-weighted
edges.  A perfect matching trivially exists and the corresponding MWPM
has weight zero.

(3) finally it is possible to find a relation between the matched
edges $M$  on $G_{\rm A}$ and a configuration of negative-weighted
loops  $\mathcal{C}$ on $G$ by  tracing the steps of the
transformation (1) back. In this regard note that each edge  contained
in $M$ that connects an additional site (square) to a duplicated  site
(circle) corresponds to an edge on $G$ that is part of a loop, see
figure \ref{fig2abcd}(d). More precisely, there are always two such edges
in $M$  that correspond to one loop segment on $G$. All the edges in
$M$ that connect like  sites (i.e.\ duplicated-duplicated, or
additional-additional)  carry zero weight and do not contribute to a
loop on $G$.  Once the set $\mathcal{C}$ of loops is found, a
depth-first search \cite{ahuja1993,opt-phys2001} can  be used to
explore $\mathcal{C}$ and  to determine the geometric properties of
the individual loops. For the weight assignment in figure
\ref{fig2abcd}(a), there is only one loop with  weight $\omega_{\mathcal
  L}\!=\!-2$ and length $\ell=8$.

Note that the result of the calculation is a collection $\mathcal{C}$
of loops such that the total loop weight, and consequently the
configurational energy  $E$, is minimized. Hence, one obtains a global
collective optimum of the system.  Obviously, all loops that
contribute to $\mathcal{C}$ possess a negative weight.  Regarding the
weight assignment in step (1) there are different possibilities  that
all result in equivalent sets of matched edges on the transformed
lattice, corresponding to the minimum-weight collection of loops on
the  original lattice. Some of these weight assignments lead to a more
symmetric  transformed graph, see e.g. \cite{ahuja1993}. However, this
is only a technical issue that does not affect the resulting loop
configuration. Albeit the  transformed graph is not symmetric, the
resulting graph (figure \ref{fig2abcd}(d))  is again symmetric.

In the following we will use the procedure outlined above to investigate the 
effect of dilution on the critical properties of NWP.

\section{Results \label{sect:results}}

So as to clarify the effect of dilution on the critical properties of
NWP, we first  need to locate the critical lines in the
$\rho-p_{\rm{I/II}}$-planes that separate the percolating from the
nonpercolating domains. Therefore, we use observables from
percolation theory and a finite-size-scaling analysis.  Besides the
weight $\omega_{\mathcal{L}}$ and length $\ell=\sum_{\{i,j\}\in
\mathcal{L}}1$ of an individual loop, we determine its linear
extensions by projecting it onto the perpendicular lattice axes. The
larger of the two is called the spanning length of the  loop
and the smaller one is called its roughness.  Consequently, a loop is
called percolating if its spanning length 
is equal to the system size $L$.  This
is a binary decision for each disorder instance that is further used
to obtain the  percolation probability $P_L(\rho,p_{\rm{I/II}})$ for a lattice of
a certain size at a given value  of the disorder and dilution
parameters $\rho$ and $p=p_{\rm{I/II}}$, respectively.  Note that system spanning
loops are a direct consequence of the periodic boundary conditions we
employ. For completeness, we note that we used the number $1$ to
signify a sample containing a  percolating loop and $0$ otherwise.  In
practice, we fix one of the parameters, say $\rho$, and tune the value
of $p$ so as to  determine the critical point $p_c$ beyond which
percolating loops emerge in the limit of large system sizes at that
particular value of $\rho$. For clarity, $P_L(\rho,p)$ at fixed $\rho$
is simply denoted as $P_L(p)$. Referring to percolation theory
\cite{stauffer1994}, $P_L(p)$ is expected to scale as $P_L(p)\sim
f_1[(p-p_c)L^{1/\nu}]$, where $\nu$ denotes the critical exponent
that describes the divergence of the typical length scale as the
critical point is approached. Since we consider string-like
objects, it is possible to find more than just one percolating loop
per disorder instance.  In a preceding work \cite{melchert2008} we
found that the average number of percolating loops per sample
satisfies the scaling relation $\langle N \rangle\sim
f_2[(p-p_c)L^{1/\nu}]$ governed by the same values for $p_c$ and $\nu$
as the percolation probability. Therein $\langle N \rangle$ was found
to exceed a value of $1$ above the critical point, similar to what was
found earlier for a different loop percolation model subject to
optimization  constraints \cite{pfeiffer2003}.  This is in contrast to
usual random-bond percolation, where the objects of interest are
cluster of connected sites on the lattice and where there is a unique
percolating cluster above the percolation threshold.

\begin{figure}[t!]
\centerline{
\includegraphics[width=1.0\linewidth]{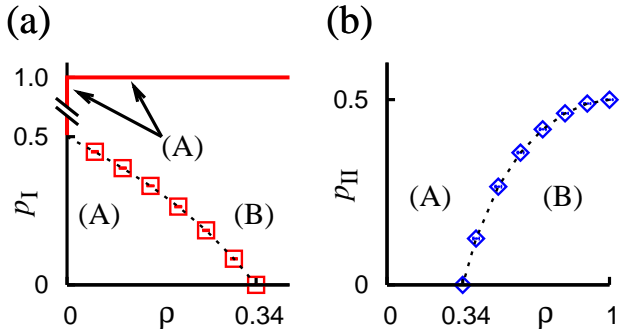}}
\caption{ Critical lines in the disorder--dilution planes.  Phase
diagram for type I and type II dilution, (a) and (b) respectively.
(A) signifies a domain in the disorder--dilution plane without
percolating loops and (B) labels a domain where percolating loops are
present.
\label{fig3ab}}
\end{figure}  

%
\begin{table}[b!]
\caption{\label{tab1}
Critical points in the disorder--dilution plane for type I dilution. 
From left to right: numerical value of the  disorder 
($\rho$) parameter and the dilution ($p_{\rm{I}}$)  along the critical line 
(the values without errorbars
where kept fixed) and critical exponent $\nu$ 
that describes the divergence of the correlation length.
$S$ measures the quality of the data collapse as the mean-square 
distance of the data sets to the master curve in units of the standard 
error.
} 
\begin{ruledtabular}
\begin{tabular}[c]{l@{\qquad}l@{\qquad}l@{\qquad}l}
$\rho$ &  $p_{\rm{I}}$    & $\nu$ & $S$ \\
\hline
0.05 & 0.449(2) & 1.49(13) & 1.19 \\    
0.10 & 0.394(2) & 1.52(14) & 1.19 \\
0.15 & 0.334(1) & 1.52(2)  & 1.43 \\
0.20 & 0.264(2) & 1.52(9)  & 1.00 \\
0.200(2) & 0.264&       1.45(12) &  0.91 \\
0.25 & 0.184(2) & 1.52(11) & 1.21 \\
0.249(2) & 0.184&       1.51(14) &  0.91 \\
0.30 & 0.088(2) & 1.51(7)  & 0.72 \\
0.340(1) & 0.000&       1.49(7)  &  0.91 \\
\end{tabular}
\end{ruledtabular}
\end{table}

\begin{figure}[t!]
\centerline{
\includegraphics[width=1.0\linewidth]{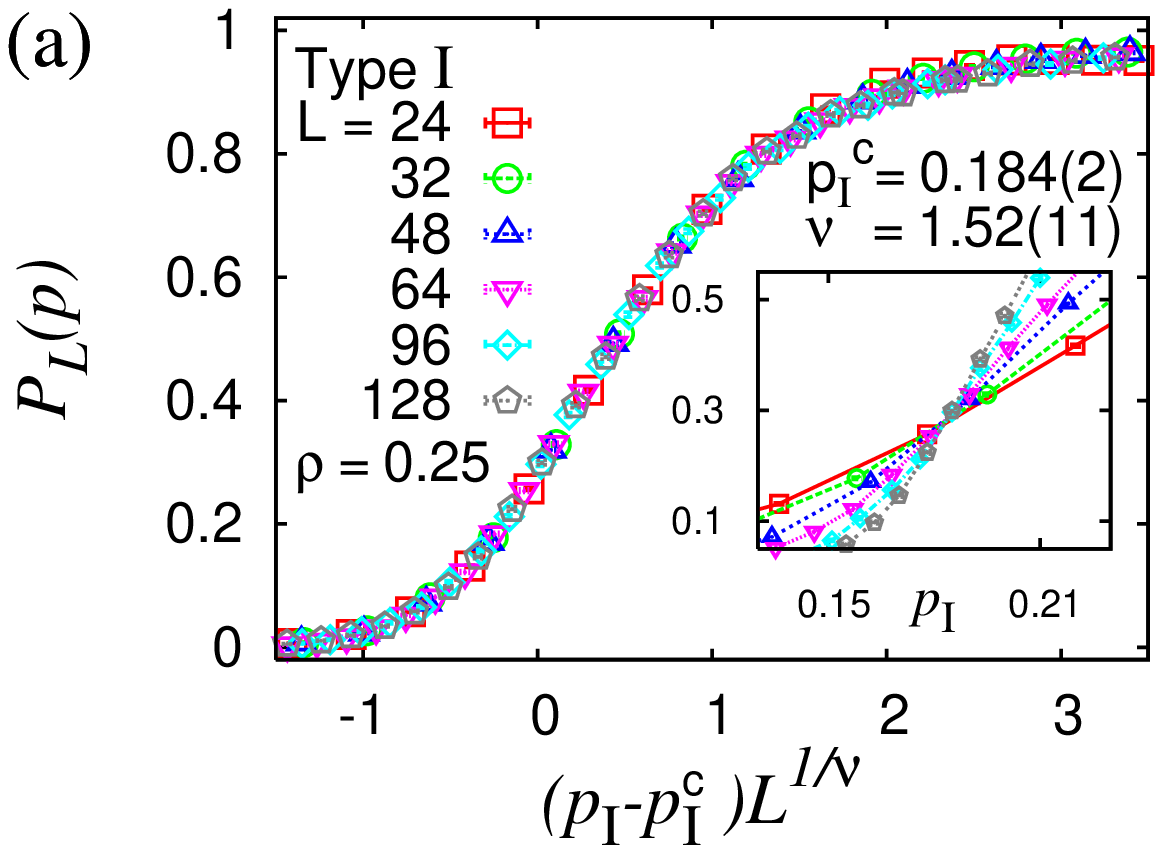}}
\centerline{
\includegraphics[width=1.0\linewidth]{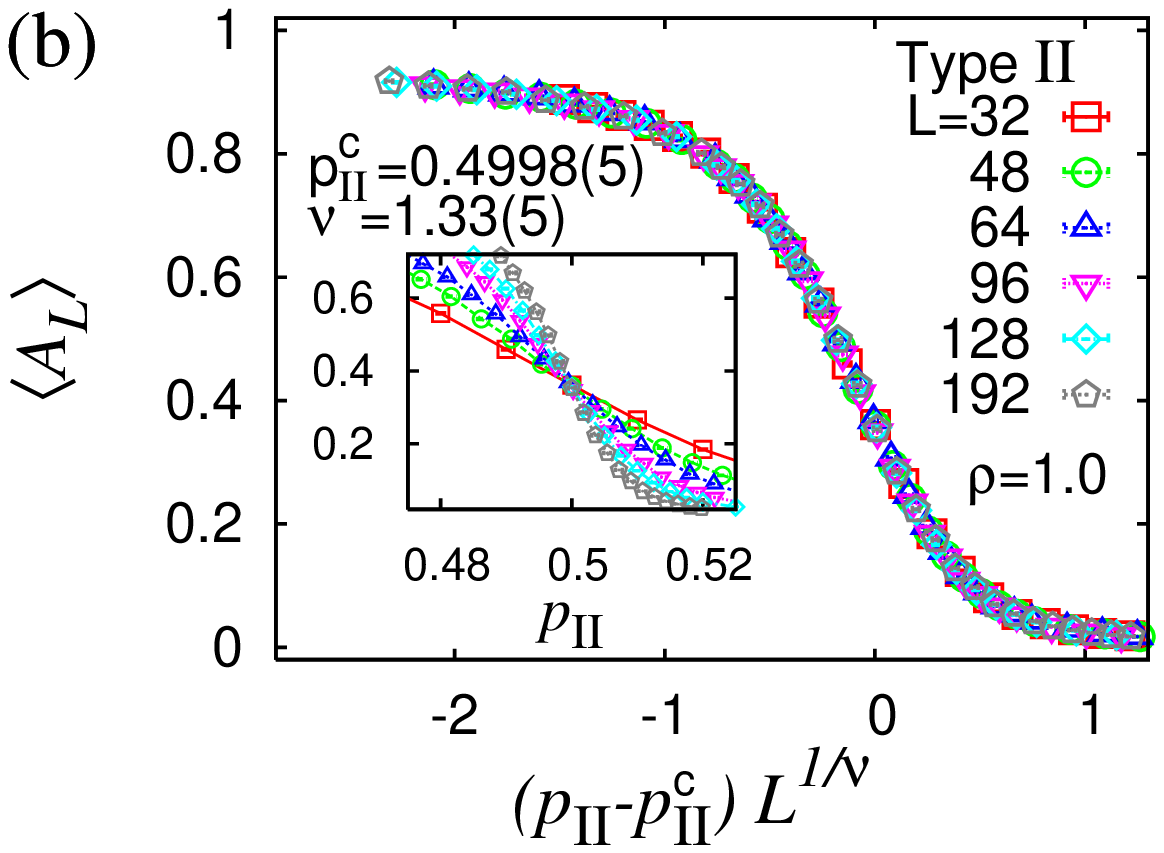}}
\caption{
Scaling analysis to probe points along the critical lines in 
the disorder--dilution planes.
(a) For a disorder parameter fixed to $\rho=0.25$, the scaling 
        analysis of the percolation probability $P_L$ yields a critical 
value $p_{\rm{I}}^{\rm{c}}=0.184(2)$ for 
        the dilution above which percolating loops appear in 
        the system. It further yields a critical exponent $\nu=1.52(11)$. 
(b) For a disorder parameter fixed to $\rho=1.0$, the scaling analysis of the
        normalized average box size $A_L$ yields 
        a critical value $p_{\rm{II}}^{\rm{c}}=0.4998(5)$ for the dilution parameter
        and a critical exponent $\nu=1.33(5)$. 
\label{fig4ab}
}
\end{figure}  

\subsection{Type I dilution}
As described above, we allow for a fraction $p_{\rm{I}}$ of edges that carry
zero weight.  A loop can include those edges without increasing its
weight or the configurational energy. As a consequence, for values of
$p_{\rm{I}}>0$ percolating  loops can emerge even if the disorder parameter
$\rho$ has a value  smaller than its critical value along the $p_{\rm{I}}=0$
axis, i.e. $\rho=0.34$.  In particular for $p_{\rm{I}}>0.5$, there is a system
spanning cluster of edges with zero weight and  for $L\rightarrow
\infty$ an arbitrarily small value of $\rho$ will suffice to generate
percolating loops on the lattice.

First, we probed the critical line that separates the
nonpercolating (small $p_{\rm{I}}$ at fixed $\rho$) and the percolating
domains in the  disorder--dilution plane. For this purpose, we
 performed simulations for
systems of size $L=24 \ldots 128$, the number of samples ranging from
$32000$ for $L=24$ to $12800$ for the largest systems.  
Within the scaling analysis, we fixed the disorder parameter $\rho$ at
values listed in table \ref{tab1} and varied the dilution $p_{\rm{I}}$ so as
to determine the  critical point $p_{\rm{I}}^{\rm{c}}$ at which percolating
loops emerge.  As an indicator for the transition, we therefore study
the finite-size-percolation  probability $P_L(p_{\rm{I}})$.  The resulting
phase diagram is depicted in figure \ref{fig3ab}(a)  and a scaling
analysis for one of the points along the critical line is shown in
figure \ref{fig4ab}(a).  A data collapse involving data sets for
different system  sizes $L$ can be used to estimate $p_{\rm{I}}^{\rm{c}}$ and
$\nu$.  The quality of the data collapse is therein measured by the
mean square  distance $S$ of the data sets to the master curve in
units of the standard  error \cite{houdayer2004}, see table
\ref{tab1}.  For the example illustrated in figure \ref{fig4ab}(a),
where the disorder parameter was  fixed to $\rho=0.25$, the scaling
analysis resulted in a value $p_c=0.184(2)$ for  the dilution above
which percolating loops are present in the system.  It further yields
a value $\nu=1.52(11)$ for the critical exponent that describes the
divergence of the correlation length.   For consistency, we checked
some of the points by performing simulations along both axis: in
addition to the data obtained at fixed $\rho$ we performed simulations
at fixed $p_{\rm{I}}$, probing the critical point along the $\rho$-axis.
For example, probing along the $p_{\rm{I}}$-axis at $\rho=0.2$ we found
$p_{\rm{I}}^{\rm{c}}=0.264(2)$, while  probing along the $\rho$-axis at
$p_{\rm{I}}=0.264$ we obtained $\rho=0.200(2)$ in complete agreement.  Albeit
the sample numbers in the present study are somewhat smaller than in
\cite{melchert2008} that led to the estimate of $\nu$ at $p_{\rm{I}}=0$,  the
numerical values we obtain for $\nu$ along the critical line are
consistent  with the value of $\nu=1.49(7)$ reported there.  This
leads us to conclude that type I dilution does not change the
universality class of NWP.  Note that it is not possible to find
negative weighted loops along the $p_{\rm{I}}=1$ and $\rho=0$ axes since at
least one edge with a negative edge weight is need in order to form a
loop.

%
\begin{table}[b!]
\caption{\label{tab2}
Critical points in the disorder--dilution plane for type II dilution. 
From left to right: numerical value of the disorder 
($\rho$) and dilution ($p_{\rm{II}}$) parameter along the critical line 
(the values without errorbars where kept fixed) and critical exponent $\nu$ 
that describes the divergence of the correlation length.
$S$ measures the quality of the data collapse as the mean square 
distance of the data sets to the master curve in units of the standard 
error.} 
\begin{ruledtabular}
\begin{tabular}[c]{l@{\qquad}l@{\qquad}l@{\qquad}l}
 $\rho$    & $p_{\rm{II}}$ & $\nu$ & $S$ \\
\hline
0.340(1) & 0.0    & 1.49(7)  &  0.91 \\
0.4 & 0.125(2)  & 1.49(9)   & 0.97 \\   
0.5 & 0.265(2)  & 1.49(15)  & 1.02 \\   
0.6 & 0.357(2)  & 1.49(11)  & 1.03 \\   
0.7 & 0.420(2)  & 1.47(11)  & 1.02 \\   
0.8 & 0.463(1)  & 1.47(11)  & 0.93 \\   
0.9 & 0.4893(9) & 1.41(8)   & 1.20 \\   
1.0 & 0.4998(5) & 1.33(5)  & 0.98 \\    
\end{tabular}
\end{ruledtabular}
\end{table}

\begin{figure}[t!]
\centerline{
\includegraphics[width=1.0\linewidth]{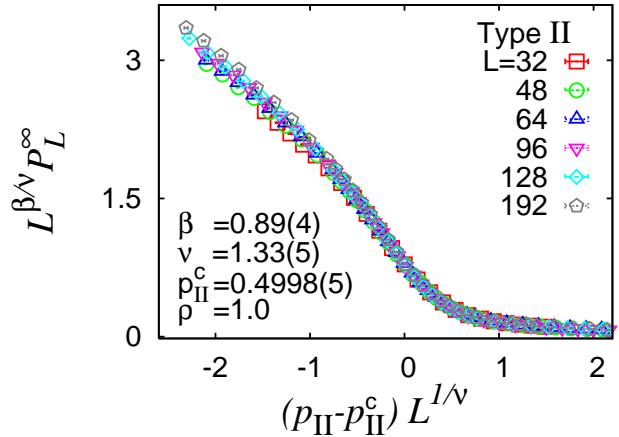}}
\caption{
FSS of the average normalized loop length $P_L^{\infty}$ that leads to the 
estimate of $\beta$ listed in table \ref{tab3}. Here, the data collapse is
best near the critical point while there are corrections to scaling off 
criticality.
\label{fig7}}
\end{figure}  

\begin{figure}[t!]
\centerline{
\includegraphics[width=1.0\linewidth]{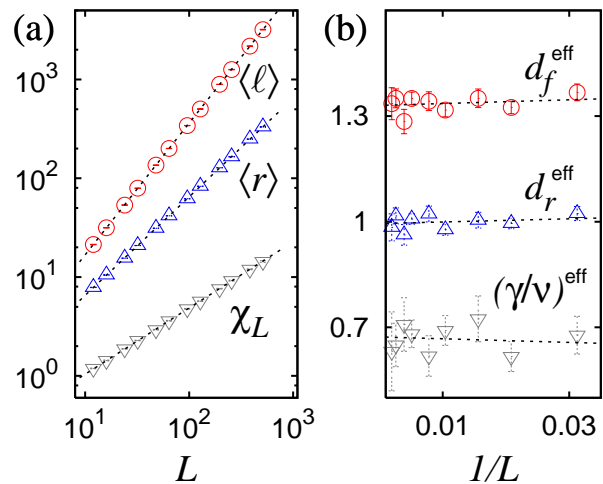}}
\caption{
Analysis of the fractal dimension $d_f$, the roughness exponent
$d_r$ and the exponent $\gamma/\nu$ that relates to the
fluctuations of the loop lengths at $\rho=1.0$ and $p_{\rm{II}}=0.4998(5)$. 
(a)
Estimation of the above exponents by means of a pure fit to the
power law data.
(b)
Estimation of the asymptotic exponents by extrapolation using the effective 
(local) exponents of the data points shown in (a).
\label{fig5ab}}
\end{figure}  
\subsection{Type II dilution}
As detailed above, we allow for a fraction $p_{\rm{II}}$ of absent edges on
the lattice.  In terms of random-bond percolation, one can expect that
the lattice decomposes into independent clusters of connected sites
for $p_{\rm{II}}>1/2$. Therefore its very unlikely to find negative
weighted loops with a spanning length equal to $L$ for values
of the dilution that exceed $p_{\rm{II}}=1/2$. Proceeding as above, we
fixed the value of the disorder parameter to the values listed in
table \ref{tab2} and tuned the value of the dilution parameter so as
to probe the corresponding critical points $p_{\rm{II}}^{\rm{c}}$,  above which
percolating loops cease to exist (large $p_{\rm{II}}$ at fixed $\rho$).  We
performed simulations for systems of size $L=32 \ldots 128$.
The number of samples ranged from
$19200$ for $L=32$ to
$12800$  for the largest
systems. We performed in  each case a FSS
analysis. Only at $\rho=1.0$ we carried out further  simulations at
$L=192$ ($9600$ samples) so as to decrease finite-size effects
furthermore and to establish a more reliable scaling analysis to
settle a value for the critical  exponent $\nu$.  Figure
\ref{fig4ab}(b) illustrates a  scaling analysis for fixed $\rho=1.0$,
corresponding to the rightmost data point in  figure
\ref{fig3ab}(b). Instead of the percolation probability shown for the
type I dilution in figure \ref{fig4ab}(a), we here show the average
area  $\langle A_L \rangle \equiv \langle \ell_x \times \ell_y\rangle
/ L^2$ of the smallest box that fits the largest loop on the lattice
normalized by the system size.  In the same manner as the percolation
probability, this quantity approaches a step  function in the limit of
large system sizes and its finite size scaling behavior follows the
analogous form $\langle A_L \rangle\sim f_3[(p_{\rm{II}}-p_{\rm{II}}^{\rm{c}})L^{1/\nu}]$.  For
fixed $\rho=1.0$ the FSS of the observable $A_L$ yields a critical
value $p_{\rm{II}}^{\rm{c}}=0.4998(5)$  of the dilution parameter and the
critical exponent $\nu=1.33(5)$ with a quality $S=1.25$ of the data
collapse.  These findings are in complete agreement with the
corresponding values listed in table \ref{tab2} that where obtained by
means of the percolation probability.  It is interesting to note that
the exponent $\nu$ steadily decreases from the value $\nu=1.49(7)$
found for the undiluted model at $\rho=0.340(1)$ to the value
$\nu=1.33(5)$ at $\rho=1.0$ and  $p_{\rm{II}}=0.4998(5)$, see table
\ref{tab3}. The latter two values are also characteristic  for random
bond percolation in 2d, that is the accompanying process for NWP
subject to the  type of dilution under investigation. The change of
the exponent $\nu$ indicates the introduction of type
II dilution changes the universality class of NWP.  
In particular, since the value of $\nu$ is compatible with the standard
percolation exponent, we ask the question whether the NWP is here
equivalent to standard percolation, although we look for loops
of negative weight instead for percolating clusters.
In the following
we present a more detailed analysis of the critical point at
$\rho=1.0$ and $p_{\rm{II}}=0.4998(5)$, which shows that indeed
the NWP behavior at $\rho=1.0$ is {\em not} equivalent to 
standard percolation.

%
\begin{table*}
\caption{\label{tab3}
Critical points and exponents.  From
left to right: Type (RBP: random bond percolation,
NWP: negative weight percolation ),
dilution ($p_{\rm{II}}$) and disorder ($\rho$) parameters 
(note that there is no entry for RBP since the disorder parameter 
$\rho$ introduced here has no analog in usual random percolation), 
critical exponent of the correlation length $\nu$,  percolation 
strength $\beta$, exponent $\gamma$, Fisher exponent $\tau$ and fractal 
dimension $d_f$ at criticality.} 
\begin{ruledtabular}
\begin{tabular}[c]{@{}llllllll}
Type         & $p_{\rm{II}}$  & $\rho_c$     & $\nu$    & $\beta$    & $\gamma$ & $\tau$  & $d_f$   \\
\hline
RBP &0.5        & --            & 4/3      & 5/36       & 43/18    & 187/91     & 91/48    \\
NWP &0.0        &  0.340(1)     & 1.49(7)  & 1.07(6)    & 0.77(7)  & 2.59(3)    & 1.266(2) \\ 
NWP &0.4998(5)  &  1.0          & 1.33(5)  & 0.89(4)    & 0.89(6)  & 2.54(5)    & 1.333(2) \\ 
\end{tabular}
\end{ruledtabular}
\end{table*}
\subsection{Detailed analysis at $\rho=1.0$ and $p_{\rm{II}}^{\rm{c}}=0.4998(5)$}
Above we found the critical point $p_{\rm{II}}^{\rm{c}}$ and the exponent $\nu$
from a FSS analysis of  the average normalized box size
$A_L$. Further,  the average normalized loop length
$P_L^{\infty}\equiv \langle \ell \rangle/L^{2}$ can  be used to obtain
a further critical exponent, namely the percolation strength $\beta$,
by use of the scaling assumption  $P_L^{\infty}\sim L^{-\beta/\nu}
f_4[(p_{\rm{II}}-p_{\rm{II}}^{\rm{c}})L^{1/\nu}]$ as suggested by the corresponding
observables known from percolation theory.  From the FSS of
$P_L^{\infty}$ we obtained the critical exponent $\beta=0.89(6)$ with
a quality $S=0.97$ of the scaling assumption, see figure \ref{fig7}
table \ref{tab3}.  In the formula above, the value of $\ell$ refers to
the length of the largest loop found for  a given realization of the
disorder.
\begin{figure}[t!]
\centerline{
\includegraphics[width=1.0\linewidth]{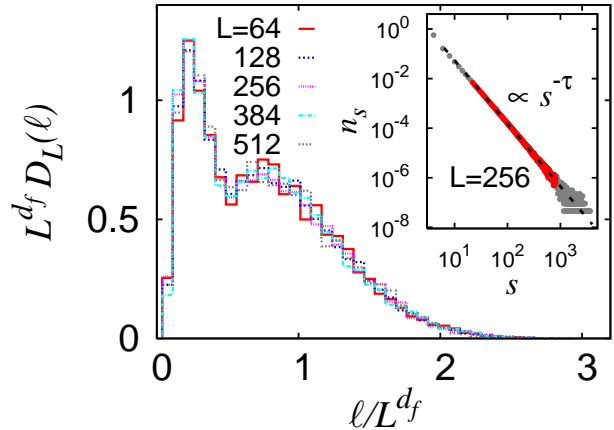}}
\caption{
Detailed analysis at $\rho=1.0$ and $p_{\rm{II}}$. 
Rescaled distribution $D_{L}(\ell)$ 
of the length $\ell$ of the largest loop found for each realization 
of the disorder.
The inset shows the 
distribution $n_s$ of all loop-lengths $s$,
excluding the spanning loops, at the critical point. Grey data points were
omitted from the fit that led to the estimate of $\tau=2.54(5)$.
\label{fig6}}
\end{figure}  
At criticality this reduces to $P_L^{\infty}\sim L^{-\beta/\nu}$. 

Next, we consider the fractal (scaling) dimension of the loops 
defined via 
$\langle \ell \rangle \sim L^{d_f}$.
 The value of $d_f$ is bounded by the
intrinsic dimension of a line and the topological dimension of the
environment to  $1\leq d_f \leq 2$. 
From scaling theory, we obtain the relation
 $d_f=2-\beta/\nu$.
This scaling relation may
further be used to check for consistency as we employ two different
methods to estimate critical exponents, i.e.\  a data collapse
involving data at different values of $p_{\rm{II}}$ and an analysis of the
algebraic scaling of different observables right at the critical
point.   A simple
fit to the power law data, shown in figure \ref{fig5ab}(a), yields the
exponent $d_f=1.333(2)$  with  quality $Q=0.51$ \cite{q_value}.
 This is compatible  with the scaling relation within error
bars, given the values
deter minded above for $\beta$ and $\nu$.
Interestingly, 
the value of $d_f$ found here coincides with the fractal dimension of
self avoiding walks on a regular  lattice ($d_f^{\rm{SAW}}=4/3$) and of
exterior perimeters measured for $2d$ percolation clusters at the
percolation threshold \cite{freund1991}.
Moreover, it compares well with the value $d_f^{\rm{BB}}\!=\!1.34(2)$ that 
describes the scaling of the average backbone perimeter with system size, 
measured for random percolation at the percolation threshold \cite{manna1989}.
This suggests, in analogy with the conclusions of \cite{manna1989}, that
negative weighted loops at $(\rho,p_{\rm{II}})\!=\!(1.0,0.4998)$ belong 
to the same universality class as self-avoiding walks.
The roughness of the loops is expected to display the critical
scaling $\langle r \rangle\sim L^{d_r}$ with a roughness exponent
$d_r=1$. The
fit to the power law data, also shown in figure \ref{fig5ab}(a), yields 
indeed the
exponent $d_r=0.998(2)$ with quality
$Q=0.33$, 
highlighting the self-similar nature of the loops.

The susceptibility like quantity $\chi_L= (\langle \ell^2\rangle
-\langle \ell \rangle^2)/L^{2}$, that measures the mean square
fluctuation of the loop length, exhibits the critical  scaling
$\chi_L\sim L^{\gamma/\nu}$ that allows to determine the critical
exponent $\gamma=0.89(4)$ with a quality $Q=0.56$ of the fit, see
again figure \ref{fig5ab}(a).  It is connected to the other exponents
via the scaling relation $\gamma +2\beta=2\nu$, which is fulfilled
within error bars.

Note that an  analysis of the
effective (local) exponents of the data belonging to $\langle \ell \rangle$,
$\langle r \rangle$ and $\chi_L$ yields the asymptotic scaling exponents
$d_f=1.33(1)$, $d_r=0.99(1)$ and $\gamma=0.89(6)$ compatible with
the values found above but with somewhat larger errorbars, see figure
\ref{fig5ab}(b).  


As evident from figure \ref{fig6}, the distributions
$D_L(\ell)$ of the largest loops $\ell$ found for each realization of
the disorder for different system sizes $L$ show a nice data collapse
after a rescaling to the form $D_L(\ell)=L^{-d_f}
f_5[\ell/L^{d_f}]$. This means that not only the average $\langle \ell
\rangle$ but the  full distribution scales with $d_f$, again
highlighting the fractal nature of the  loops. Therein, the peak close
to $\ell/L^{d_f}=0.25$ stems from the non-spanning loops and the peak
close to $\ell/L^{d_f}=0.75$ stems from those that truly span the
lattice along at least one direction.  At the critical point, we
further expect the distribution $n_s$ of all loop lengths $s$,
excluding loops with a spanning length $L$, to  exhibit
an algebraic scaling $n_s \sim s^{-\tau}$, where $\tau$ signifies the
Fisher exponent. The exponent $\tau$ characterizes the complete
ensemble of loops and  it is related to the fractal exponent of the
loops via $\tau-1=d/d_f$. Here, we found  a value of $\tau=2.54(5)$,
see inset of figure \ref{fig6}, in agreement with the latter  scaling
relation.


\section{Conclusions \label{sect:conclusions}}
In summary, we performed numerical simulations in order to probe the
effect of dilution on the critical properties of NWP on $2d$ lattice
graphs.  In this context and by means of exact combinatorial
optimization algorithms we have investigated two different types of
dilution.  One type affecting the distribution of weights that are
associated with the edges,  the other type
having an impact on the topology
of the underlying lattice.  We used observables from percolation
theory and a FSS analysis so as to probe critical
points along the critical lines in the disorder-dilution plane that
separate domains that allow/disallow system spanning,
i.e.\ percolating, loops.  We found that the first type of dilution
does not change the universality class of NWP, i.e.\ the critical
exponents are the same as for the undiluted model.  The second type of
dilution very well leads to a change of the universality class: As the
disorder on the lattice increases, the critical exponent $\nu$  as
well as the critical point of dilution approach the corresponding
values  known from usual random bond percolation,
although the process, also due to its string-like nature,
is clearly different from standard percolation, also visible
through the values of the other critical exponents. 
A more detailed
analysis of the critical exponents for one particular critical point
in the disorder-dilution plane, i.e.\ $(\rho,p_{\rm{II}})\!=\!(1.0,0.4998)$,
verified that the exponents are connected by the usual  scaling relations.
Moreover, the results for the scaling behavior of the percolating loops
suggest that negative weighted loops at that particular critical 
point belong to the same universality class as self-avoiding walks.


\begin{acknowledgments}
LA  acknowledges a scholarship of the German academic exchange service DAAD
within the ``Research Internships in Science and Engineering'' (RISE) program
and the City College Fellowship program for further support.
We further acknowledge financial support from the VolkswagenStiftung (Germany)
within the program ``Nachwuchsgruppen an Universit\"aten''. The
simulations were performed at the  ``Gesellschaft f\"ur
Wissenschaftliche Datenverarbeitung'' in G\"ottingen
(Germany), and the GOLEM I cluster for scientific computing at the
University of Oldenburg (Germany).
\end{acknowledgments}


\bibliographystyle{plain}
\bibliography{nwp_diluted.bib}

\begin{thebibliography}{10}
\bibitem{comment_cookrohe}
For the calculation of minimum-weighted perfect matchings we use Cook and Rohes
  blossom4 extension to the Concorde library.
\end{thebibliography}

\end{document}